\documentclass{elsarticle}
\usepackage{gensymb}
\usepackage{graphics}
\usepackage{amsmath}
\usepackage{amsfonts}
\usepackage{color}
\usepackage{natbib}
\usepackage{hyperref}
\usepackage{lineno}
\usepackage{bbm}
\usepackage[normalem]{ulem}
\usepackage{tikz}

\begin{document}

\title{Modeling the Second Harmonic in Surface Water Waves Using Generalizations of NLS}
\author[1]{Hannah Potgieter}
\author[1]{John D. Carter$^*$}
\author[2]{Diane M. Henderson}

\cortext[cor1]{Corresponding author, carterj1@seattleu.edu}
\address[1]{Mathematics Department, Seattle University, Seattle, WA, 98122, USA}
\address[2]{Mathematics Department, Penn State University, State College, PA, 16801, USA}

\begin{abstract}
  If a wavemaker at one end of a water-wave tank oscillates with a particular frequency, time series of downstream surface waves typically include that frequency along with its harmonics (integer multiples of the original frequency).  This behavior is common for the propagation of weakly nonlinear waves with a narrow band of frequencies centered around the dominant frequency such as in the evolution of ocean swell, pulse propagation in optical fibers, and Bose-Einstein condensates.  Presented herein are measurements of the amplitudes of the first and second harmonic bands from four surface water wave laboratory experiments.  

  \smallskip

  The Stokes expansion for small-amplitude surface water waves provides predictions for the amplitudes of the second and higher harmonics given the amplitude of the first harmonic.  Similarly, the derivations of the NLS equation and its generalizations (models for the evolution of weakly nonlinear, narrow-banded waves) provide predictions for the second and third harmonic bands given amplitudes of the first harmonic band.  We test the accuracy of these predictions by making two types of comparisons with experimental measurements.  First, we consider the evolution of the second harmonic band while neglecting all other harmonic bands.  Second, we use explicit Stokes and generalized NLS formulas to predict the evolution of the second harmonic band using the first harmonic data as input.  Comparisons of both types show reasonable agreement, though predictions obtained from dissipative generalizations of NLS consistently outperform the conservative ones.  Finally, we show that the predictions obtained from these two methods are qualitatively different.

\end{abstract}

\maketitle

\section{Introduction}

When a mechanical wavemaker at one end of a water-wave tank oscillates with a single frequency, $\omega_0$, time series of downstream surface waves show the dominant frequency, $\omega_0$, along with its harmonics, $2\omega_0$, $3\omega_0$, etc.  The knowledge of the existence of harmonics in small-amplitude surface waves goes back at least to the work of Stokes~\cite{Stokes}.  See Flick \& Guza~\cite{Flick} for a more detailed discussion and relevance to the generation of laboratory waves.  Data from ocean field experiments showed that the second harmonic is important in the generation of rogue waves~\cite{dias16}.  Second harmonic measurements have been made in a wide variety of real-world wave phenomena.  In pulse propagation in optical fibers, the generation of second harmonics was first observed by Sasaki \& Ohmori~\cite{OpticalFibers}.  In Bose-Einstein condensates, second harmonics have been experimentally observed, see for example Anderson et al.~\cite{BEC}.  In this paper, we study the evolution of the second harmonics in the water-waves setting.

For small-amplitude waves, the classical Stokes expansion~\cite{Stokes} provides predictions for the amplitudes of the second and higher harmonics given the amplitude of the first harmonic.  By comparison, instead of assuming that the leading-order part of the solution to the water-wave problem is composed of a single sinusoid as in the Stokes expansion, the derivation of the nonlinear Schr\"odinger (NLS) equation assumes that the leading-order part of the solution is composed of a sum of sinusoids from a narrow band of frequencies.  NLS is derived from the water-wave problem by an asymptotic expansion in both wave steepness and spectral bandwidth, which are assumed to be of the same order.  The NLS derivation provides formulas that predict the evolution of the waves in the second and third harmonic bands given amplitudes of the waves in the first harmonic band.  These formulas generalize the Stokes predictions.  See Zakharov~\cite{Zakharov} for the original derivation of the NLS equation as a model of gravity waves on deep water and Johnson~\cite{Johnson} for a more modern derivation.  The derivations of many generalized NLS equations, including the Dysthe, dissipative NLS, and viscous Dysthe equations, provide an asymptotic formula for the evolution of the second harmonic band in terms of first harmonic band, see equation (\ref{eqn:NLSB2}) below.  To leading order, the formulas from these NLS-type models are the same as the prediction obtained from the NLS equation.

There have been many comparisons between mathematical models and experimental measurements of the first harmonic band; see for example Lake \& Yuen~\cite{LakeYuen}, Lake et al.~\cite{LakeEtAl}, Lo \& Mei~\cite{Lo}, Trulsen et al.~\cite{TrulsenExpt}, Segur et al.~\cite{Segur}, Wu et al.~\cite{Wu}, Ma et al.~\cite{MaDong}, Simanesew et al.~\cite{Simanesew}, and Carter et al.~\cite{Butterfield}.  The evolution of the second harmonic has been studied in much less detail, especially in the NLS regime.  Lake and Yuen~\cite{LakeYuen} were interested in the growth of sidebands, the modes with frequencies nearby the dominant one.  They considered an explanation for discrepancies between measured and predicted growth rates of the sidebands to be due to the disagreement between the measured second harmonic amplitude and that predicted by the Stokes expansion.   However, Crawford et al.~\cite{Crawford} stated ``...this effect is far less significant than was believed and should be disregarded.''  Thus, we examine the interplay between the first and second harmonic bands in experiments.

In particular, we examine the evolution of the first and second harmonic bands in four series of water wave laboratory experiments.  We make comparisons of two types.  First is a ``direct'' comparison, see Section \ref{SectionDirectComp}, in which we consider the evolution of the second harmonic band as if it were the dominant band while neglecting the first harmonic band.  (The evolution of the first harmonic band while ignoring the other harmonic bands, for the data considered herein, was previously examined in Segur et al.~\cite{Segur}, Carter \& Govan~\cite{Govan}, and Carter et al.~\cite{Butterfield}.)  For this comparison, we take measurements of the second harmonic band at the first experimental gauge, use them as initial conditions for numerical simulations of NLS and its generalizations, and compare the numerical predictions with the measurements from the downstream gauges.  We refer to comparisons of this type as ``direct'' since second harmonic band data at the first gauge is used to predict second harmonic band data at the downstream gauges.  

Second are two ``indirect'' comparisons in which we take measurements of the first harmonic band from the first gauge, use them as initial conditions for numerical simulations of the envelope equations, and use asymptotic formulas to obtain predictions for the second harmonic band at the downstream gauges.  We refer to comparisons of this type as ``indirect'' because data from the first harmonic band from the first gauge is used to predict second harmonic band data at the downstream gauges.  The first indirect comparison, see Section \ref{SMComparisons}, is a single-mode comparison based on the Stokes expansion, see formula (\ref{eqn:Stokesb2}).  The second indirect comparison, see Section \ref{section:BandComp}, is a band comparison based on the formula that is obtained in the derivation of the NLS-type equations, see formula (\ref{eqn:NLSB2}).  The main goal of this work is to determine how accurately NLS and its generalizations model the evolution of the second harmonic band in water waves.  To our knowledge, this work is the first explicit test on the accuracy of NLS and its generalizations as models of the second harmonic.

Although the main goal of this work is to determine the validity of NLS-type predictions for the evolution of the second harmonic band, it also touches on the free wave versus forced wave question.  In a laboratory experiment in which a wavemaker oscillates at frequency $\omega_0$, the water surface is forced to oscillate primarily at the frequency $\omega_0$.  The corresponding wavenumber, $k_0$, is determined by the deep-water dispersion relation
\begin{equation}
  \omega_0^2=(gk_0+\Gamma k_0^3),
  \label{LDR}
\end{equation} 
where $g$ and $\Gamma$ represent the acceleration due to gravity and the coefficient of kinematic surface tension respectively.  Flick \& Guza~\cite{Flick} discussed how a mechanical wavemaker is a dynamical system so that when it oscillates at a prescribed frequency, it will also put some nonzero energy in the harmonics no matter how precise the control system.  If the wavemaker puts in non-zero mechanical energy at the $n$th harmonic frequency, $n\omega_0$, then there will also be waves with frequency, $n\omega_0$.  These ``free'' waves have wavenumbers that are determined by the dispersion relation.  If there were no mechanical motion at frequency $n\omega_0$, then the $n$th harmonic would exist only through weak nonlinearity as a forced wave and its wavenumber would be $n k_0$, where $k_0$ is the wavenumber of the dominant mode.  These forced harmonics would then have the same phase speed as the dominant mode, would be phase-locked to the dominant mode, and would not satisfy the dispersion relation.  See for example, Dean \& Dalrymple \cite{DD} (pp.~170-185) for a discussion of wavemaker theory in general and free and forced waves in particular.

The remainder of this paper is organized as follows.  Section \ref{SectionModels} presents the six models considered herein and summarizes their derivations from the governing system.  The models include three conservative and three dissipative equations.  The conservative models are the Stokes expansion, the NLS equation, and the Dysthe equation.  The dissipative models are the dissipative NLS, the viscous Dysthe, and the dissipative Gramstad-Trulsen equations.  Section \ref{SectionExperiments} contains a description of the four experiments considered and tables listing the experimental parameter values.  Section \ref{SectionDirectComp} presents the results from the direct comparisons.  There, we first show comparisons of the measured, narrow band of amplitudes of the dominant (first-harmonic) band with predictions of the evolution obtained from numerical computations of the five NLS-type models.  These results are summarized in Table \ref{table:direct} and Figure \ref{fig:har1}.  Then we show comparisons of the measured narrow-band of amplitudes centered at the second harmonic with predictions of its evolution obtained from numerical computations of the five NLS-type models.  The results are shown in Table \ref{table:direct} and Figure \ref{fig:har2}.  Section  \ref{SectionIndirectComp} contains the results from the indirect comparisons with results summarized in Tables \ref{table:asymExpt1} and \ref{table:asymExpt2}. The single-mode indirect comparisons are discussed in Section \ref{SMComparisons} with results shown in Figure \ref{fig:Stokes}. The band indirect comparison is discussed in Section \ref{section:BandComp}.  Finally, Section \ref{SectionSummary} contains a summary and discussion of the results.

\section{Model Equations}
\label{SectionModels}

Consider the following system for the motion of an infinitely-deep, weakly-dissipative, two-dimensional fluid
\begin{subequations}
\label{eqn:Wusystem}
\begin{equation}
 \phi_{xx}+\phi_{zz} = 0,~~~  \text{ for }  - \infty < z < \eta, \label{eqn:Wua}
\end{equation}
\begin{equation}
 \phi_{t} +\frac{1}{2}|{\nabla\phi}|^{2}+g\eta -\Gamma\frac{\eta_{xx}}{(1+\eta_x^2)^{3/2}}= -\alpha \phi_{zz} ,\: ~~~ \text{at}  \: z = \eta, \label{eqn:Wub}
\end{equation}
\begin{equation}
 \eta_{t}+\eta_{x}\phi_{x} = \phi_{z},~~~\text{ at } z = \eta, \label{eqn:Wuc}
\end{equation}
\begin{equation}
 |{\nabla\phi}| \to 0,~~~ \text{ as } z \to -\infty. \label{eqn:Wud}
\end{equation}
\end{subequations}
Here $\phi = \phi(x, z, t)$ represents the velocity potential of the fluid, $\eta = \eta(x, t)$ represents the free-surface displacement, 
$x$ is the horizontal coordinate,
 $z$ is the vertical coordinate, $t$ is the temporal coordinate, $g$ represents the acceleration due to gravity, $\Gamma$ is the coefficient of kinematic surface tension, and $\alpha\ge 0$ is a constant such that $\alpha \phi_{zz}$ represents dissipation from all sources.  In particular, $\alpha=2C_g\delta_e/k_0^2$, where $C_g$ is the linear group velocity, $\delta_e$ is the spatial decay rate (typically measured in experiments), and $k_0$ is the wavenumber.  See Wu et al.~\cite{Wu} for a more detailed discussion of this form for dissipation.  The classical boundary-value problem for conservative water waves is obtained from this system by setting $\alpha=0$.  The $\alpha=0$ case is colloquially known as the ``water-wave problem.''

\subsection{Stokes expansion}

In order to find a small-amplitude asymptotic solution to (\ref{eqn:Wusystem}) with $\alpha=\Gamma=0$, Stokes~\cite{Stokes} assumed that the surface displacement has the form
\begin{equation}
  \eta (x, t) = \epsilon b \: \mbox{e}^{i(\omega_0 t - k_0 x)} + \epsilon^2 b_2 \: \mbox{e}^{2i(\omega_0 t -  k_0 x)} + ... + c.c.,
\label{eqn:StokesEta}
\end{equation}
where $\epsilon = 2 a_0 k_0 \ll 1$ is the dimensionless wave steepness, $k_0>0$, $\omega_0$, and $a_0$ are constants representing the wavenumber, frequency, and amplitude of the dominant wave respectively, and $c.c.$ stands for complex conjugate.  We assume that $\omega_0>0$ because the experiments considered herein are unidirectional.  Although Stokes~\cite{Stokes} did not, we allow for surface tension such that the frequency and wavenumber are related by the dispersion relation given in (\ref{LDR}),  see for example, Harrison~\cite{Harrison} and Akers \& Nicholls~\cite{AkersNicholls}.  Then the ansatz ({\ref{eqn:StokesEta}}), leads to the following expansions for the amplitudes of the first and second harmonics
\begin{subequations}
  \begin{equation}
    b=a_0\left(1-\epsilon^2\frac{8g^2+31gk_0^2\Gamma+14k_0^4\Gamma^2}{16(g-2k_0^2\Gamma)(g+k_0^2\Gamma)}+\mathcal{O}(\epsilon^3)\right),
  \end{equation}
  \begin{equation}
    b_2=\frac{k_0(g+k_0^2\Gamma)}{g-2k_0^2\Gamma}a_0^2+\mathcal{O}(\epsilon).
    \label{eqn:Stokesb2}
  \end{equation}
\end{subequations}
The indirect comparisons use measurements and computations of $b$ from the NLS-type models to compute $b_2$ using equation (\ref{eqn:Stokesb2}).  We examine the accuracy of these formulas in Sections \ref{SectionDirectComp} and \ref{SectionIndirectComp}.

\subsection{NLS and its generalizations}

One way to generalize the idea behind the Stokes expansion is to allow the leading-order coefficient to comprise a narrow band of frequencies instead of a single frequency.  See Ablowitz \& Segur~\cite{AS} and references relevant here discussed below.  Here we follow the work of Carter \& Govan~\cite{Govan} and assume that the surface displacement has the form
\begin{equation}
\label{eqn:etaexpansion}
    \eta (x, t, X, T) = \epsilon B(X,T) \: \mbox{e}^{i(\omega_0 t - k_0 x)} + \epsilon^2 B_2(X, T) \: \mbox{e}^{2i(\omega_0 t -  k_0 x)} + ... + c.c..
\end{equation}
Here the coefficients of the harmonics depend on the slow variables, $X = \epsilon x$ and $T = \epsilon t$, instead of being constants as in the Stokes expansion.  Similarly to the Stokes expansion, the coefficients $B$ and $B_i$ for $i = 2, 3, 4, ...$ have their own asymptotic expansions.  We assume that dissipative effects are small by setting $\alpha = \epsilon^2 \bar{\alpha}$.  The frequency and wavenumber are related by the dispersion relation given in equation (\ref{LDR}).  Carrying out the asymptotics to fourth order in $\epsilon$ and then using the nondimensionalization defined by
\begin{subequations}
\label{eqn:COV}
\begin{align}
  \xi &= \omega_{0} T  - \frac{2\omega_0^2}{g+3k_0^2\Gamma} X,\label{eqn:COVa}\\
  \chi &= \epsilon k_0 X,\label{eqn:COVb}\\
  \bar{B} (\xi, \chi) &= k_0 B (x, t),\label{eqn:COVc}\\
  \gamma&=\frac{k_0^2}{g}\Gamma,\\
  \delta &=\frac{4k_0^2}{\omega_0}\frac{1+\gamma}{1+3\gamma}\bar{\alpha},\label{eqn:COVd}
\end{align}
\end{subequations}
leads to the dimensionless viscous Dysthe (vDysthe) equation (see Dysthe~\cite{Dysthe} for the derivation of the original inviscid Dysthe equation)
\begin{equation}
\label{eqn:vDysthe}
    iB_{\chi}+\mu_1B_{\xi\xi}+\mu_2|B|^{2}B + i\delta B+\epsilon\bigg(i\mu_3B_{\xi\xi\xi}+i\mu_4B^{2}B_{\xi}^{*}+i\mu_5|{B}|^{2}B_{\xi}+\mu_6\big{(}{\mathcal{H}}(|{B}|^{2})\big{)}_{\xi}B+\delta\mu_7 B_{\xi}\bigg) = 0,
\end{equation}
where the bars have been dropped for convenience.  Here $\xi$ represents nondimensional time, $\chi$ represents nondimensional distance down the tank, $B$ represents the nondimensional complex amplitude of the envelope, $\gamma$ represents nondimensional surface tension, $\delta$ represents nondimensional dissipation, the $\mu_j$ are the real coefficients defined by
\begin{subequations}
  \begin{equation}
  \mu_1=\frac{(1+\gamma)(1-6\gamma-3\gamma^2)}{(1+3\gamma)^3},
  \end{equation}
  \begin{equation}
    \mu_2=\frac{8+\gamma+2\gamma^2}{2+2\gamma-12\gamma^2},
  \end{equation}
  \begin{equation}
    \mu_3=\frac{4\gamma(1+\gamma)^2(5-10\gamma-3\gamma^2)}{(1+3\gamma)^5},
  \end{equation}
  \begin{equation}
    \mu_4=\frac{(1+\gamma)(1-3\gamma)(8+\gamma+2\gamma^2)}{(-1+2\gamma)(1+3\gamma)^3},
  \end{equation}
  \begin{equation}
    \mu_5=\frac{2(1+\gamma)(-16+13\gamma-55\gamma^2+12\gamma^4)}{(1-2\gamma)^2(1+3\gamma)^3},
  \end{equation}
  \begin{equation}
    \mu_6=-\frac{16(1+\gamma)^3}{(1+3\gamma)^3},
  \end{equation}
  \begin{equation}
    \mu_7=\frac{5+10\gamma+9\gamma^2}{(1+3\gamma)^2},
  \end{equation}
\end{subequations}
and $\mathcal{H}$ is the Hilbert transform defined by
\begin{equation}
\label{eqn:Hilbert}
    {\mathcal{H}}\big(f(\xi)\big) = \sum_{k = - \infty}^{\infty} -i ~ \text{sgn}(k) \hat{f}(k) \mbox{e}^{2\pi i k \xi/L},
\end{equation}
where $L$ is the $\xi$-period of the experimental time series and the Fourier transform of a function $f(x)$ is defined by
\begin{equation}
\label{eqn:Ftrans}
    \hat{f}(k) = \frac{1}{L} \int_0^L f(\xi) \mbox{e}^{- 2\pi i k \xi/L} d\xi.
\end{equation}
Equation (\ref{eqn:vDysthe}) generalizes the vDysthe equation derived by Carter \& Govan~\cite{Govan} to include surface tension.  In the derivation, one also finds the following relationship between the evolution of the second and first harmonic bands (presented in dimensional form for ease of experimental comparison)
\begin{equation}
  B_2=\frac{k_0(g+k_0^2\Gamma)}{g-2k_0^2\Gamma}B^2+i\epsilon\frac{g^2+5gk_0^2\Gamma -2k_0^4\Gamma^2}{(g-2k_0^2\Gamma)^2}BB_X+\mathcal{O}(\epsilon^2).
  \label{eqn:NLSB2}
\end{equation}
 Note the similarity between this equation and the Stokes result given in equation (\ref{eqn:Stokesb2}).  Though some of the generalized NLS models we consider are dissipative and others are conservative, the relationship between $B_2$ and $B$ given in equation (\ref{eqn:NLSB2}) is the same to leading order.  Lake \& Yuen~\cite{LakeYuen} emphasized the importance of this relationship in their study of the Benjamin-Feir instability.  In the indirect comparison of measured and predicted values of $B_2$, we use (i) experimentally measured values of $B$ in (11) and (ii) computations of $B$ in (11) obtained from the NLS-type evolution models. These comparisons are based upon information from the narrow-band of frequencies centered at the first harmonic.

The NLS equation,
\begin{equation}
\label{eqn:NLS}
    iB_{\chi}+\mu_1B_{\xi\xi}+\mu_2|{B}|^{2}B = 0,
\end{equation}
is obtained from the vDysthe equation by setting $\delta=\epsilon=0$.  Zakharov~\cite{Zakharov} first derived the NLS equation as a model of gravity waves (without surface tension) on deep water in the late 1960s.  Djordjevic \& Redekopp~\cite{DjordjevicRedekopp} derived the NLS equation for gravity waves with surface tension on water of finite depth.  The NLS equation has been well studied; see for example Ablowitz \& Segur~\cite{AS} (includes a derivation that allows for surface tension) and Sulem \& Sulem~\cite{Sulem}.  It also arises as an approximate model for a wide range of other physical phenomena including pulse propagation along optical fibers~\cite{opticswaveguides}, Langmuir waves in a plasma~\cite{Pecseli}, and superfluids such as Bose-Einstein condensates~\cite{Pitaevskii,Gross}.  The water wave results presented below may be generalizable to these other phenomena.

Other well-known generalizations of the NLS equation are found by examining various limits of the vDysthe equation.  When $\epsilon = 0$ and $\delta>0$, the vDysthe equation reduces to the dissipative NLS (dNLS) equation
\begin{equation}
\label{eqn:dNLS}
    iB_{\chi}+\mu_1B_{\xi\xi}+\mu_2|{B}|^{2}B + i\delta B= 0.
\end{equation}
When $\delta = 0$ and $\epsilon>0$, the vDysthe equation reduces to the Dysthe~\cite{Dysthe} equation
\begin{equation}
\label{eqn:Dysthe}
    iB_{\chi}+\mu_1B_{\xi\xi}+\mu_2|{B}|^{2}B + \epsilon\bigg(i\mu_3B_{\xi\xi\xi}+i\mu_4B^{2}B_{\xi}^{*}+i\mu_5|{B}|^{2}B_{\xi}+\mu_6\big{(}{\cal H}(|{B}|^2)\big{)}_{\xi}B\bigg) = 0.
\end{equation}
Additionally, generalizing the work of Gramstad \& Trulsen~\cite{dGT}, Carter et al.~\cite{Butterfield} proposed the following generalization of the NLS equation
\begin{equation}
\label{eqn:dGT}
    iB_{\chi}+\mu_1B_{\xi\xi}+\mu_2|{B}|^{2}B + i\delta B+\epsilon\bigg(i\mu_3B_{\xi\xi\xi}+i\mu_4|{B}|^{2}B_{\xi}+\mu_6\big{(}{\cal H}(|{B}|^2)\big{)}_{\xi}B+\delta\mu_7 B_{\xi}\bigg)+i\epsilon^{2}\delta\mu_8 B_{\xi\xi} = 0,
\end{equation}
where
\begin{equation}
  \mu_8=\frac{2(1+\gamma)(-5+10\gamma+3\gamma^2)}{(1+3\gamma)^3}.
\end{equation}
The addition of the $\mathcal{O}(\epsilon^2)$ term comes from extending the derivation of the vDysthe equation one order and removes a flaw in the vDysthe model, see \cite{Butterfield} for details.  We refer to this equation as the dissipative Gramstad-Trulsen (dGT) equation.

Numerical solutions of all the envelope equation models were obtained using the high-order operator splitting methods introduced by Yoshida~\cite{yoshida}.  Periodic boundary conditions in $\xi$ were imposed.  The linear parts of the envelope equations were solved exactly in Fourier space using fast Fourier transforms and the nonlinear parts were either solved exactly (NLS, dNLS) or using fourth-order Runge-Kutta (Dysthe, vDysthe, dGT).

\section{Experiments}
\label{SectionExperiments}

We examine four experiments conducted in the William G.~Pritchard Fluid Mechanics Laboratory in the Mathematics Department at Penn State.  The first two of these experiments were first studied in Segur et al.~\cite{Segur} and the second two were first studied in Carter et al.~\cite{Butterfield}.  Following the nomenclature in \cite{Butterfield}, we refer to the four experiments as Experiments A, B, C, and D.  The wave channel used for Experiments A and B was 13.1 m long.  The wave channel used for Experiments C and D was 15.2 m long.  Both channels had glass bottoms and sidewalls and were 0.254 m wide.  The tank walls were cleaned with alcohol and then water was added.  The air-water interface was cleaned by skimming (A and B) or blowing (C and D) the interfacial layer to one end of the tank where it was vacuumed.  The resulting still-water depth for all experiments was 0.20 m. Waves were generated in all four experiments with anodized, wedge-shaped plungers that spanned the width of the tank and were oscillated vertically using feed-back, programmable control.  The cross-section of the wedge was exponential for A and B (with a fall-off that corresponds to the velocity field for a 3.33 Hz wave) and was triangular for C and D (with a slope corresponding to a linear approximation to the exponential of the paddle for A and B). For Experiments A and B, which used PMAC - Delta Tau Data Systems for motion control, the wedge was oscillated with a time series given by 
\begin{equation}
    \eta_p(t)=a_f\sin(\omega_{0f} t)\big{(}1+r\sin(\omega_{1f}t)\big{)},
\end{equation}
where $a_{f}$ is the forcing amplitude of the first harmonic, $\omega_{0f}$ is the frequency of the first harmonic, $r$ is the ratio of perturbation amplitude to $a_f$, and $\omega_{1f}$ is the perturbation frequency.  Experiments C and D used ARCS software for motion control.  For C, the wedge was oscillated with a time series given by
\begin{equation}
  \eta_p(t)=a_f\sin(\omega_{0f} t)\big{(}1-r\cos(\omega_{1f}t)\big{)},
\end{equation}
and for Experiment D,
\begin{equation}
    \eta_p(t)=a_f\sin(\omega_{0f} t)+a_{f}r\sin\big{(}(\omega_{0f}+\omega_{1f})t\big{)}.
\end{equation}
The forcing for Experiment D was chosen so that only the upper sideband was forced.  Note that in all four experiments, only the first harmonic and its first upper and lower sidebands were forced.  The prescribed forcing motion did not seed the harmonics.  However, all mechanical (and physical) systems naturally force the higher harmonics.  In this case, the feedback control minimized harmonics in the mechanical motions.  Table \ref{table:WaveMakerParams} contains the wavemaker parameters used in each of the four experiments.  Time series of surface displacement were recorded at 11-13 locations down the tank.  The sampling frequency for the time series collected in Experiments A and B was 350 samples/sec while the sampling frequency for Experiments C and D was 500 samples/sec.  A more detailed description of the procedures for Experiments A and B is included in Segur et al.~\cite{Segur}.

\begin{table}
\begin{center}
\begin{tabular}{ |c|c|c|c|c| } 
\hline
Parameter & Expt A & Expt B & Expt C & Expt D \\
\hline
$\omega_{0f}/(2\pi) \:\:\: ({\text{Hz}})$ & $3.336$ & $3.333$ & $3.333$ & $3.333$ \\ 
$\omega_{1f}/(2\pi) \:\:\: ({\text{Hz}})$ & $0.17$ & $0.17$ & $0.11$ & $0.11$ \\
$a_{f} \:\:\: ({\text{m}})$ & $2.5*10^{-3}$ & $2.5*10^{-3}$ & $5.0*10^{-3}$ & $5.0*10^{-3}$\\
$r$ & $0.14$ & $0.33$ & $0.50$ & $0.50$\\
\hline
\end{tabular}
\caption{Wavemaker parameters for the four experiments.}
\label{table:WaveMakerParams}
\end{center}
\end{table}

\begin{figure}
  \centering
      \includegraphics[width=16cm]{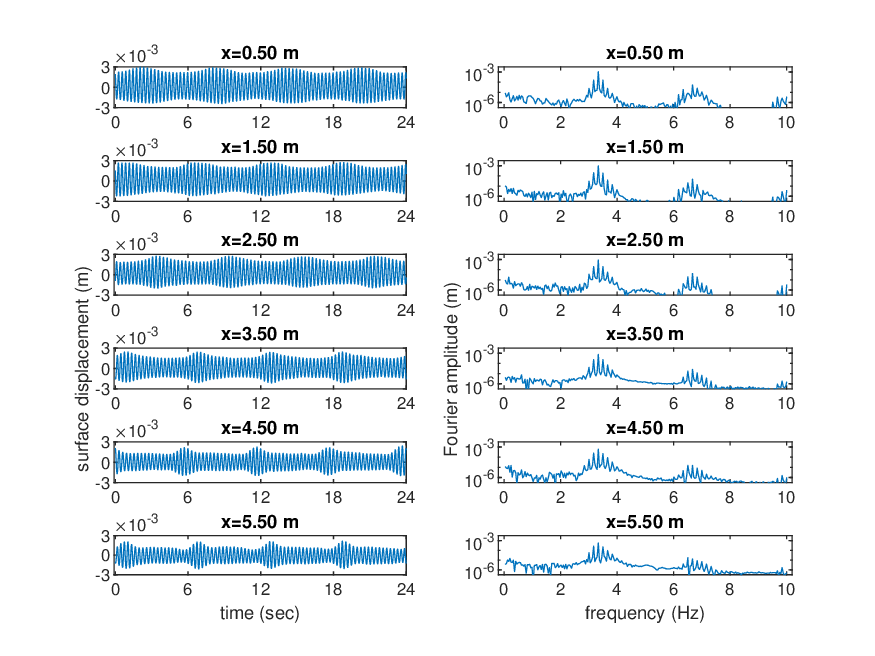}  
      \caption{Plots from every other gauge for Experiment A. The left column contains plots of surface displacement (in m) versus time (in sec). The right column contains log-scale plots of the magnitudes of the corresponding Fourier coefficients (in m) versus frequency (in Hz).}
      \label{fig:data}
  \end{figure}

Each of the four experiments consisted of a set of 11-13 sub-experiments with gauges located $1.28+0.50(m-1)$ meters for $m = 1, 2, . . . , M$ from the wavemaker.  For the purposes of this paper, we define $x=0$ to be the location of the gauge closest to the wavemaker.  The values of $M$ for the four experiments are included in Table \ref{table:ExptParams}.  Figure \ref{fig:data} contains plots of the surface displacement versus time and the corresponding Fourier magnitudes versus frequency for Experiment A.  For conciseness, only the time series from every other gauge are shown.  The dominant clusters in the time series from the first gauge are used to define the harmonic bands.  The first and second harmonic bands are defined by the intervals $[1.67, 5.00]$ Hz and $[5.00, 8.33]$ Hz respectively.  We refer to the first harmonic as the $3.33$ Hz peak although for some experiments this value rounds to $3.34$ Hz. Similarly, we refer to the second harmonic as the $6.66$ Hz peak.  We do not examine the the third or higher harmonic bands because their measured amplitudes are too small. 

\begin{table}
  \begin{center}
  \begin{tabular}{ |c|c|c|c|c| } 
  \hline
  Parameter & Expt A & Expt B & Expt C & Expt D \\
  \hline
  $M$ & $12$ & $11$ & $13$ & $13$ \\ 
  $N$ & $41$ & $39$ & $50$ & $45$\\
  $t_{f} \:\:\: ({\text{sec}})$ & $24.28$ & $23.40$ & $30.00$ & $27.00$\\
  $\omega_{0}/(2\pi) \:\:\: ({\text{Hz}})$ & $3.336$ & $3.333$ & $3.333$ & $3.333$ \\ 
  $\epsilon$ & $9.541*10^{-2}$ & $9.256*10^{-2}$ & $6.277*10^{-2}$ & $6.781*10^{-2}$\\
  $\delta$ & $0.2755$ & $0.3362$ & $0.5238$ & $0.8480$\\
  $\Tilde{\omega}_{0}/(2\pi) \:\:\: ({\text{Hz}})$ & $6.671$ & $6.666$ & $6.667$ & $6.667$ \\ 
  $\Tilde{\epsilon}$ & $1.605*10^{-2}$ & $1.356*10^{-2}$ & $1.377*10^{-2}$ & $1.861*10^{-2}$\\
  $\Tilde{\delta}$ & $5.123$ & $7.230$ & $5.059$ & $1.798$\\
  \hline
  \end{tabular}
  \caption{Experimentally measured parameters for each of the four experiments.  The parameters without tildes correspond to the first harmonic and the parameters with tildes correspond to the second harmonic.}
  \label{table:ExptParams}
  \end{center}
\end{table}

Table \ref{table:ExptParams} contains the experimentally measured parameters including: length of the time series, $t_{f}$; wave steepness, $\epsilon$; and dissipation parameter, $\delta$.  There are two versions of each of these parameters.  The ones without tildes were measured based on the first harmonic band, while the ones with tildes were measured based on the second harmonic band.  More specifically, $\epsilon=2a_0k_0$ where $a_0$ and $k_0$ are the amplitude and wavenumber of the 3.33 Hz wave respectively, while $\tilde{\epsilon}=2\tilde{a}_0\tilde{k}_0$ where $\tilde{a}_0$ and $\tilde{k}_0$ are the amplitude and wavenumber of the 6.66 Hz wave.  The values of the parameters without tildes are slightly different than those presented in ~\cite{Govan,Butterfield} because those studies neglected surface tension effects in the calculation of $k_0$, while here we allow for surface tension.

In particular, the values for the wavenumbers were determined using the deep-water linear dispersion relationship given in equation (\ref{LDR}) using $g=9.81$ m/sec$^2$ and $\Gamma=0.717$ m$^3$/sec$^2$.  The value for $\Gamma$ is the average of the surface tension values measured during Experiments C and D at the (also vacuumed) air-water interface of water (from the same source as for the tank) in a beaker next to the tank, using the Du No\"uy ring method.  It is close to the ``book'' value, $\Gamma=0.729$ m$^3$/sec$^2$, for the surface tension of water at $20\degree$C, see Pallas~\cite{SurfaceTension}.  It is especially important to include surface tension effects for the $6.66$~Hz wave.  Note that surface tension was not included in any of \cite{Segur,Govan,Butterfield} where only the first harmonic was examined in detail.  The value of $k_0d$ (where $d=0.20$ m is the depth of the undisturbed water) for all four experiments is $8.82$.  Since this value is well above 1, the experiments are in the deep-water regime.  

In all four experiments, the wave amplitudes decrease nearly exponentially due to dissipative effects.  This causes the value of
\begin{equation}
  \mathcal{M}(\chi)= \frac{1}{L}\int_0^L |B|^2d\xi,
  \label{eqn:mass}
\end{equation}
to decrease in a nearly exponentially manner as $\chi$ increases (i.e.~as the waves travel down the tank).  We compute $\mathcal{M}$ or $\tilde{\mathcal{M}}$ using either the first or second harmonic band, depending on which case is under consideration.  The best exponential fit of the form $\mathcal{M}(\chi)=\mathcal{M}(0)\exp(-2\delta\chi)$ for the first band leads to the $\delta$ values that are included in Table \ref{table:ExptParams}.  The best exponential fit of the form $\tilde{\mathcal{M}}(\tilde{\chi})=\tilde{\mathcal{M}}(0)\exp(-2\tilde{\delta}\tilde{\chi})$ for the second band leads to the $\tilde{\delta}$ values that are included in Table \ref{table:ExptParams}.  Note that the dimensionless variable $\tilde{\chi}$ is different than the dimensionless variable $\chi$ because it depends on $\tilde{\epsilon}$ and $\tilde{k_0}$ instead of $\epsilon$ and $k_0$, see equation (\ref{eqn:COVb}).  The values of $\tilde{\delta}$ are larger than the values of $\delta$ by factors of $18.6$, $21.5$, $9.7$, and $2.1$ for Experiments A-D respectively.  These factors vary significantly between the experiments and are all different than the factor of $4$ predicted if the dissipation was proportional to the wave number squared; see for example Young et al.~\cite{YoungBabanin}.  We note that Dore's~\cite{dore78} model for dissipation rates of 3.3 Hz and 6.6 Hz linear free waves at a clean air-water interface are $\delta_{3.3}=0.2866$ and $\delta_{6.6}=38.91$.  The measured value for the first harmonic agrees reasonably well with the predicted value, while the measured value for the second harmonic is much lower than predicted for a free, 6.6 Hz wave.  We emphasize that our empirical definitions of $\delta$ and $\tilde{\delta}$ combine all dissipative effects, regardless of their source, into a single term for each band.  

\begin{figure}
\centering
\includegraphics[width=10cm]{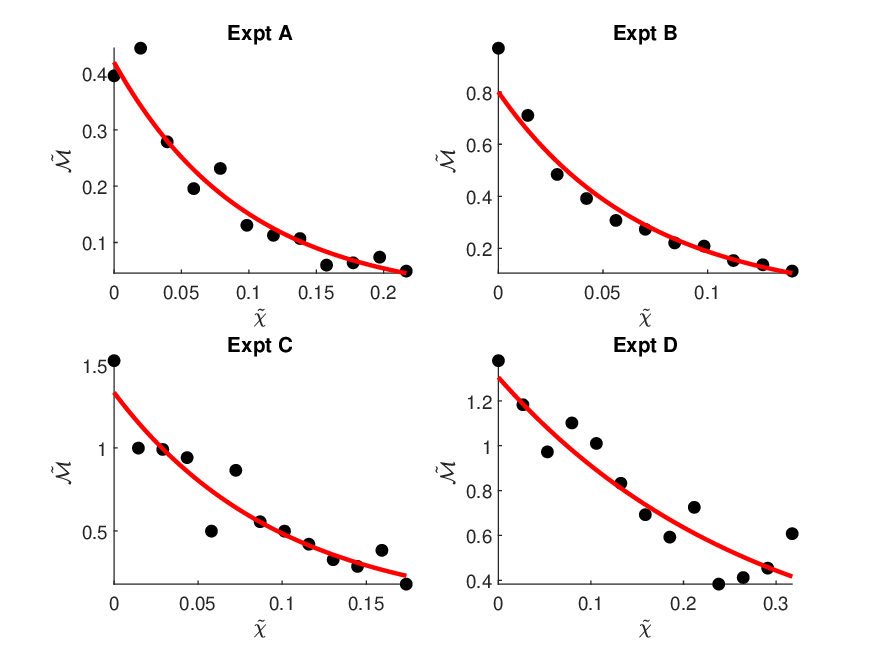}
\caption{Plots of $\tilde{\mathcal{M}}$ versus $\tilde{\chi}$  for each of the four experiments. The dots correspond to experimental measurements and the curves are the best-fit exponentials using empirically determined values for $\Tilde{\delta}$.}
\label{fig:deltas}
\end{figure}

Figure \ref{fig:deltas} includes plots of $\tilde{\mathcal{M}}$ versus $\tilde{\chi}$ for each experiment along with the best exponential fits.  Although there is strong agreement between the data and the exponential fits, the agreement between the first harmonic data and its exponential fit is even better.  See Figures 2 and 4 of \cite{Govan} and Figure 4 of \cite{Butterfield} for those plots.  Even though it is not explicitly stated in those papers, $\mathcal{M}$ was computed using only the first harmonic band.

The total energy at the first gauge is determined by computing $\mathcal{M}_{tot}(\chi = 0)$ using {\emph{all}} frequencies observed at the first gauge.  The energy in the first harmonic band is obtained by computing $\mathcal{M}(\chi = 0)$ using only the first harmonic band.  Similarly, the energy in the second harmonic band is obtained by computing $\mathcal{\tilde{M}}(\tilde{\chi} = 0)$ using only the second harmonic band.  The calculations establish that the first harmonic band constitutes close to 50\% of the total energy, whereas the second harmonic band accounts for between 0.1 and 2\% of the total energy depending on the experiment.  The remaining energy is spread across the other approximately 100 harmonic bands in a more or less random manner, though the third and fourth harmonic bands have slightly more energy than the background average.  This emphasizes the dominance of the first harmonic band in the composition of the waves.  Yet, the second harmonic band plays a non-negligible role.

Finally, we note that the higher frequency data values are noisier, particularly in the sidebands.  For this reason, we only examine the second harmonic band instead of examining multiple higher harmonic bands.

\section{Direct Comparisons}
\label{SectionDirectComp}

In this section, we compare: (i) the experimental data with the model predictions for the first harmonic band and (ii) the experimental data with the model predictions for the second harmonic band.  We call these ``direct'' comparisons because we use NLS-type equations (\ref{eqn:vDysthe}), (\ref{eqn:NLS}), (\ref{eqn:dNLS}), (\ref{eqn:Dysthe}), and (\ref{eqn:dGT}) to directly predict the experimental measurements at the downstream gauges given the experimental data at the first gauge.  The initial conditions for the first harmonic band simulations were
\begin{equation}
\label{eqn:IChar1}
    B (\xi, \chi = 0) =  \frac{k_0}{\epsilon} \sum_{n = -N}^{N} a_n \exp{\Big{(}in \frac{2\pi}{\epsilon \omega_0 t_f} \xi\Big{)}},
\end{equation}
where $a_n$ is the $n$th Fourier coefficient of the time series recorded at the first gauge and $N$ is the number of positive Fourier modes included.  We used only frequencies in the first harmonic band (i.e.~$[1.67,5.00]$~Hz) to define the initial conditions for the first harmonic simulations.  Refer to Table \ref{table:ExptParams} for the values of the parameters $\omega_0$, $k_0$, $\epsilon$, $t_f$, and $N$ for each of the four experiments.  As discussed in Section \ref{SectionExperiments}, we used $g=9.81$ m/sec$^2$ and $\Gamma=0.717$ m$^3$/sec$^2$ in all of our calculations.

Similarly, the initial conditions for the second harmonic band simulations were
\begin{equation}
\label{eqn:IChar2}
    B_2 (\Tilde{\xi}, \Tilde{\chi} = 0) = \frac{\Tilde{k}_0}{\Tilde{\epsilon}} \sum_{n = -N}^{N} \tilde{a}_{n} \exp{(in \frac{2\pi}{\Tilde{\epsilon} \Tilde{\omega}_0 t_f} \Tilde{\xi})},
\end{equation}
where only frequencies from the second harmonic band (i.e.~$[5.00,8.33]$~Hz) were used.  See Table \ref{table:ExptParams} for the values of $\tilde{\omega}_0$, $\tilde{k}_0$, $\tilde{\epsilon}$, $t_f$, and $N$ for each of the four experiments.  

\begin{figure}
\centering
\begin{minipage}[b]{0.48\linewidth}
  \includegraphics[scale=0.6]{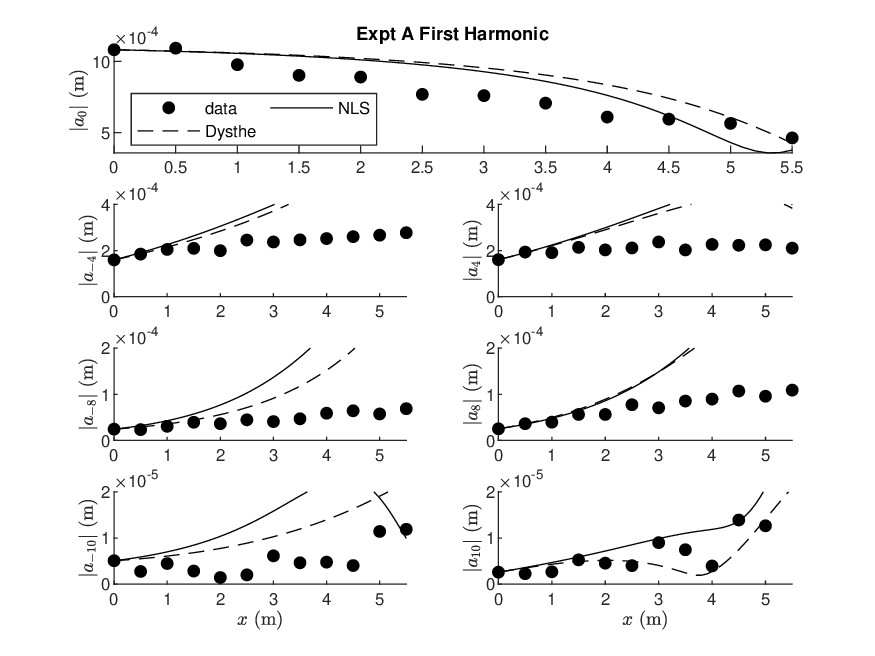}
  \label{fig:consA1}
\end{minipage}
\quad
\begin{minipage}[b]{0.48\linewidth}
  \includegraphics[scale=0.6]{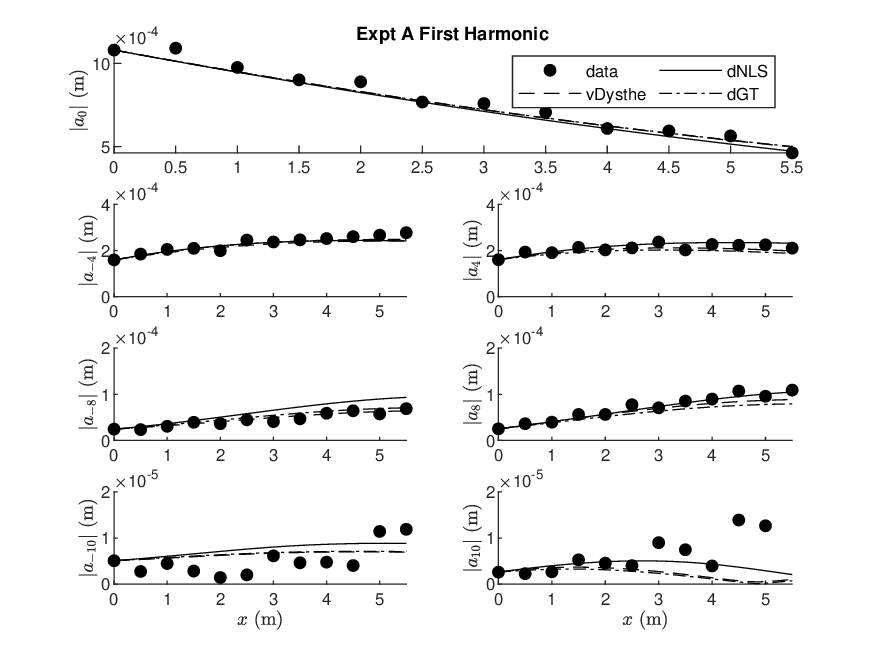}
  \label{fig:dispA1}
\end{minipage}
\quad
\caption{Plots of the magnitude of the Fourier amplitudes for the first harmonic (3.33~Hz) (top) and three of its most dominant sideband pairs (bottom) versus distance down the tank.  The dots represent experimental measurements and the curves represent the envelope equation predictions.  The conservative model predictions are on the left and the dissipative model predictions are on the right.} 
\label{fig:har1}
\end{figure}

Figure \ref{fig:har1} presents comparisons between the envelope equation predictions and the experimental measurements for the 3.33 Hz wave and six of its sidebands for Experiment A.  Only the first harmonic band was included in these comparisons and associated computations.  These comparisons include surface tension effects, while all previous comparisons of this experimental data, including those in \cite{Segur,Govan,Butterfield}, neglected surface tension.  The plots in Figure \ref{fig:har1} show that the dissipative models (dNLS, vDysthe, dGT) are more accurate than the conservative models (NLS, Dysthe).  The results for the 3.33 Hz wave in the other three experiments are similar.

\begin{figure}
\centering
\begin{minipage}[b]{0.48\linewidth}
  \includegraphics[scale=0.6]{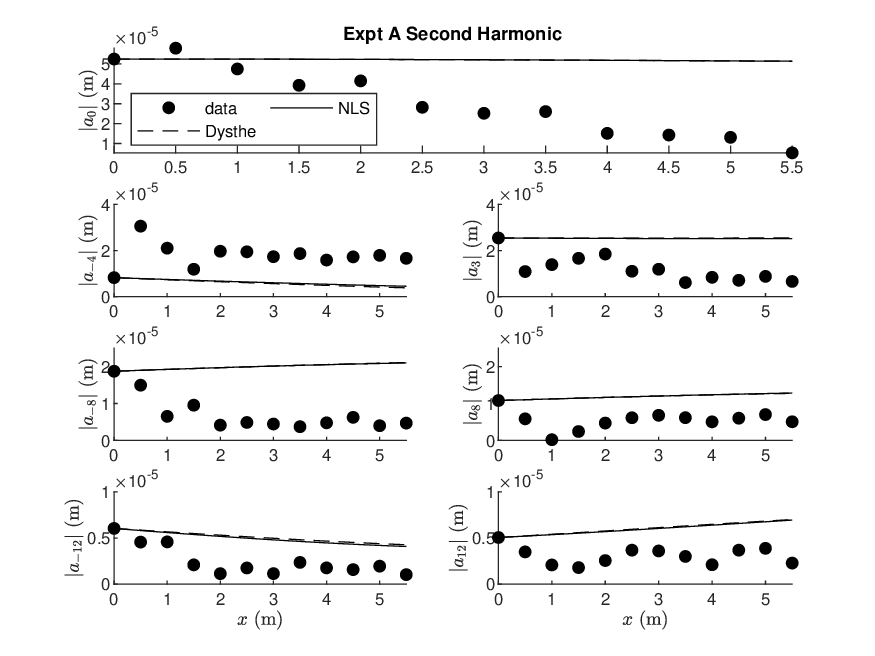}
  \label{fig:consA2}
\end{minipage}
\quad
\begin{minipage}[b]{0.48\linewidth}
  \includegraphics[scale=0.6]{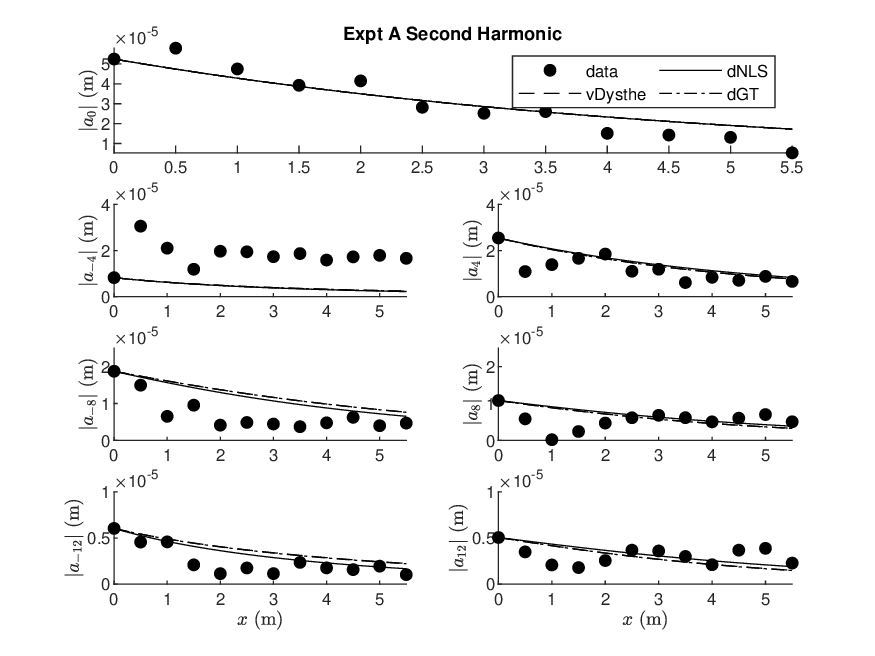}
  \label{fig:dispA2}
\end{minipage}
\quad
\begin{minipage}[b]{0.48\linewidth}
  \includegraphics[scale=0.6]{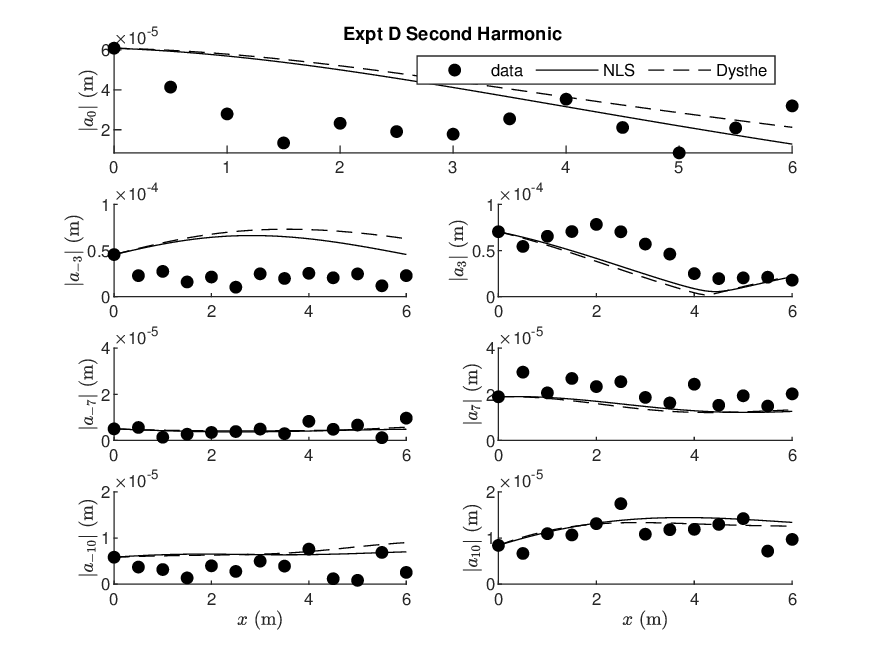}
\end{minipage}
\quad
\begin{minipage}[b]{0.48\linewidth}
  \includegraphics[scale=0.6]{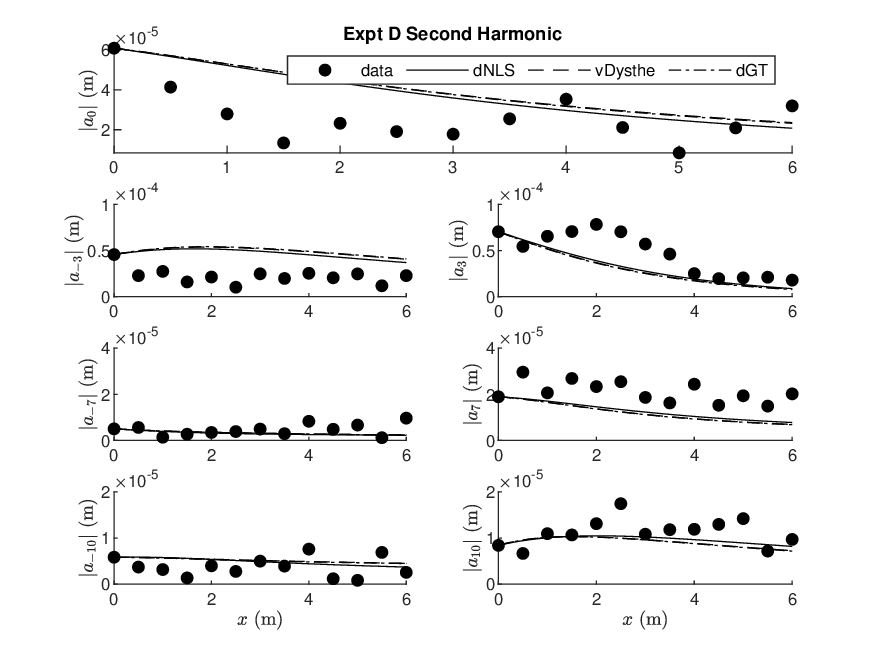}
\end{minipage}
\caption{Plots of the magnitude of the Fourier amplitudes for the second harmonic (6.66~Hz) and three of its most dominant sideband pairs versus distance down the tank for Experiments A and D. The dots represent experimental measurements and the curves represent the envelope equation predictions from the direct comparison.  The conservative model predictions are on the left and the dissipative model predictions are on the right.}
\label{fig:har2}
\end{figure}

Figure \ref{fig:har2} presents comparisons between the envelope equation model predictions and the experimental measurements for the 6.66 Hz wave and its six most dominant sidebands for Experiments A and D.  Only the second harmonic band was included in these comparisons and associated computations.  All other harmonic bands were neglected.  These are new comparisons.  The plots show that the dissipative models outperform the conservative models.  Additionally, for a given experiment, the predictions obtained from the conservative models are similar and the predictions obtained from the dissipative models are also similar.  Other than in Experiment D, there was noticeably more agreement between the two conservative models for the second harmonic band than there was for the first harmonic band predictions, although the non-conservative models provide the most accurate predictions in all experiments.  Finally, note that the sideband amplitudes plotted for each experiment vary because we choose the sidebands with the largest amplitude for each experiment.

We quantitatively compare the model predictions with the experimental time series by the following dimensionless weighted error norm to get percent errors
\begin{equation}
    \mathcal{E} = \frac{1}{(2J+1)(M-1)} \sum_{m = 2}^{M}
    \frac{ \sum_{n = -J}^{J} \big{|}|a_{n, m}^{expt}|-|a_{n, m}^{model}|\big{|}^{2}}{\mathcal{M}_m} * 100\%,
\label{eqn:error}
\end{equation}
where $J$ is the number of upper sidebands included in the comparison, M is the number of gauges used in the experiment, $\mathcal{M}_m$ represents the value of $\mathcal{M}$ (for comparisons of the first harmonic band) or $\tilde{\mathcal{M}}$ (for comparisons of the second harmonic band) at the $m$th gauge, and $a_{n, m}$ represents the amplitude of the $n$th sideband at the $m$th gauge.  We use $J=0$ for comparisons based on the Stokes prediction (i.e.~the single-mode comparisons) and $J=N$ for comparisons based on the NLS-type predictions (i.e.~the band comparisons).  This error norm calculates the average of the percent errors at each wave gauge beyond the first gauge where the initial conditions are defined.  Note that this is a different error norm than was used in \cite{Govan,Butterfield}.

Table \ref{table:direct} shows the percent errors for the comparisons discussed in this section.  The errors in the first harmonic band predictions are smaller than the errors in the second harmonic band predictions.  The three dissipative models are significantly more accurate than the conservative models for all four experiments for both bands.  For a given experiment, the predictions obtained from the dissipative models are all similar.  This is especially true for the second harmonic band predictions.  We note that neglecting surface tension, i.e.~setting $\Gamma=0$, (results not shown) slightly improves some predictions while slightly worsening other predictions in a way that does not follow an obvious pattern.  Overall, Table \ref{table:direct} and Figures \ref{fig:har1} and \ref{fig:har2} show that the direct comparisons using the dissipative models are reasonably accurate.  

\begin{table}
\centering
\begin{tabular}{ |c|c|c|c|c| } 
\hline
$B$ & Expt A & Expt B & Expt C & Expt D \\
\hline
NLS & $4.93*10^{-1}$ & $9.67*10^{-1}$ & $4.18*10^{-1}$ & $1.46*10^{0}$ \\
Dysthe & $3.72*10^{-1}$ & $8.33*10^{-1}$ & $2.86*10^{-1}$ & $1.43*10^{0}$ \\ 
dNLS & $8.46*10^{-3}$ & $3.08*10^{-1}$ & $6.18*10^{-2}$ & $8.86*10^{-2}$ \\ 
vDysthe & $7.30*10^{-3}$ & $2.03*10^{-1}$ & $5.97*10^{-2}$ & $9.22*10^{-2}$ \\ 
dGT & $8.47*10^{-3}$ & $1.85*10^{-1}$ & $5.84*10^{-2}$ & $9.39*10^{-2}$ \\ 
\hline
\end{tabular}
\begin{tabular}{ |c|c|c|c|c| } 
\hline
$B_2$ & Expt A & Expt B & Expt C & Expt D \\
\hline
NLS & $2.27*10^{0}$ & $2.50*10^{0}$ & $1.50*10^{0}$ & $8.72*10^{-1}$ \\
Dysthe & $2.28*10^{0}$ & $2.48*10^{0}$ & $1.52*10^{0}$ & $9.08*10^{-1}$ \\ 
dNLS & $3.98*10^{-1}$ & $4.75*10^{-1}$ & $3.83*10^{-1}$ & $3.79*10^{-1}$ \\ 
vDysthe & $3.97*10^{-1}$ & $5.72*10^{-1}$ & $3.93*10^{-1}$ & $3.97*10^{-1}$ \\ 
dGT & $3.99*10^{-1}$ & $5.62*10^{-1}$ & $3.94*10^{-1}$ & $4.02*10^{-1}$ \\ 
\hline
\end{tabular}
\caption{Percent error norms, $\mathcal{E}$, using (\ref{eqn:error}) for direct comparisons between the experimental data and the model predictions for first harmonic band (top) and second harmonic band (bottom).}
\label{table:direct}
\end{table}

\medskip

\subsection{Frequency Downshift}

Frequency downshift (FD) occurs when a measure of the frequency decreases monotonically as the waves travel down the tank.  Frequency upshift occurs when a measure of the frequency increases monotonically.  The two most common measures of frequency for experiments of this type are the spectral peak, $\omega_p$, and the spectral mean

\begin{equation}
    \omega_m = \frac{\mathcal{P}}{\mathcal{M}}.
\label{eqn:omegaM}
\end{equation}
Here, $\mathcal{P}$ is defined by
\begin{equation}
  \mathcal{P}(\chi)= \frac{i}{2 L}\int_0^L \left( B B_{\xi}^* -B_{\xi} B^* \right) d\xi,
  \label{eqn:P}
\end{equation}
and $\mathcal{M}$ is defined in equation (\ref{eqn:mass}).  Segur et al.~\cite{Segur} observed that FD often occurs in the second harmonic band before it occurs in the first harmonic band.  See Carter et al.~\cite{Butterfield} for a detailed discussion of frequency downshift and an examination of the accuracy of the models in predicting FD in the first harmonic band.  See \ref{AppendixA} for plots of the spectral peaks and means for both the first and second harmonic bands for all four experiments.  For the second harmonic band, the plots show that the models provide FD predictions of inconsistent accuracy in both the spectral peak and spectral mean senses.  This is unsurprising given that the models are not perfect in modeling FD in the first harmonic band which contains a significantly higher percentage of the total energy.  There is need for further work on frequency downshifting.

\section{Indirect Comparisons}
\label{SectionIndirectComp}

This section explores the relationship between first and second harmonic bands.  We call these ``indirect'' comparisons because we model the second harmonic data using first harmonic data in predictions (\ref{eqn:Stokesb2}) and (\ref{eqn:NLSB2}).  Our goal is to determine the validity of the Stokes prediction, (\ref{eqn:Stokesb2}), and the NLS prediction, (\ref{eqn:NLSB2}), for the four experiments described above.  

\subsection{Single-mode comparisons}
\label{SMComparisons}
In this section, we test the validity of the relation between $b_2$ and $b$ given in equation (\ref{eqn:Stokesb2}) (i.e.~the single-mode or Stokes relation) from a variety of perspectives.  Although the experiments included multiple frequencies in each harmonic band, in this subsection we focus on the dominant frequency in each band while ignoring all sidebands.

\begin{figure}
  \includegraphics[width=16cm]{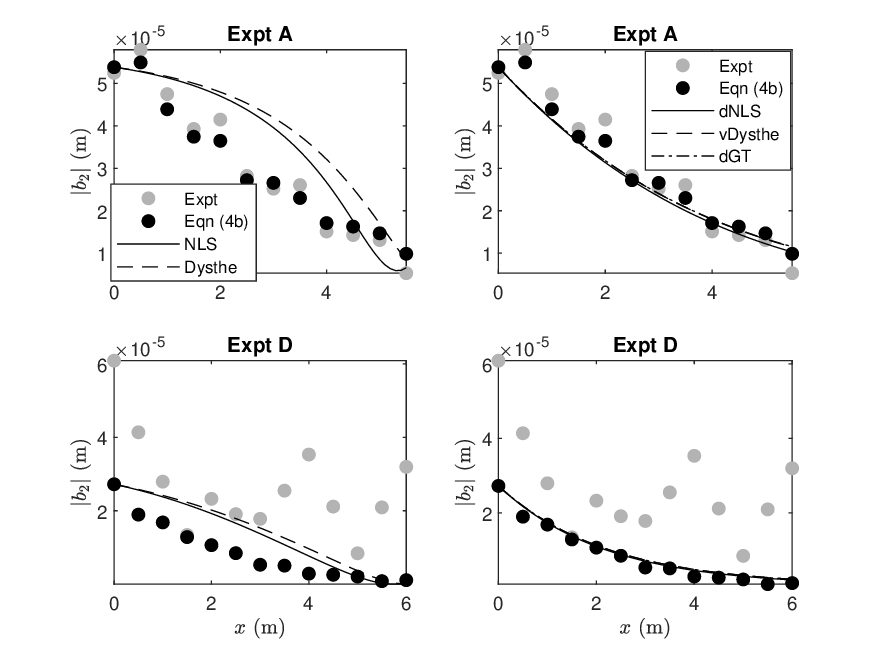}
\caption{Plots comparing the experimental 6.66 Hz data (gray dots) and the single-mode predictions for the 6.66 Hz wave.  The black dots are the predictions for the 6.66 Hz data obtained by applying (\ref{eqn:Stokesb2}) to the experimental 3.33 Hz data.  The curves are the predictions for the 6.66 Hz data obtained by applying (\ref{eqn:Stokesb2}) to the 3.33 Hz mode of the model simulations based on the first harmonic data.}
\label{fig:Stokes}
\end{figure}

Figure \ref{fig:Stokes} shows plots of the experimental data (Experiments A and D only for conciseness) for the 6.66 Hz wave (gray dots) along with the predictions obtained using the 3.33 Hz wave data and equation (\ref{eqn:Stokesb2}) (black dots).  For Experiment A, the single-mode prediction lines up well with the experimental measurements.  For Experiment D, the single-mode predictions significantly under-predict the experimental measurements.  
The results for Experiment B are roughly the same as those in Experiment A, while the results for Experiment C are roughly the same as those in Experiment D.  These results suggest that the single-mode relation given in (\ref{eqn:Stokesb2}) when applied to the experimental data is often, but not always, reasonably accurate.  

Figure \ref{fig:Stokes} also contains comparisons of the experimentally measured values of the 6.66 Hz wave (gray dots) and continuous predictions for the 6.66 Hz wave (curves).  The continuous predictions were obtained by a two-step process.  First, we solved the envelope equations using the first harmonic band measured at the first gauge as initial conditions.  Then, we extracted the 3.33 Hz mode from the simulations and applied equation (\ref{eqn:Stokesb2}) to obtain a prediction for the amplitude of the 6.66 Hz wave.  (Although the envelope equations model the entire first harmonic band, we extracted the 3.33 Hz wave while neglecting the sidebands for the comparisons in this subsection.  Comparisons of the entire band are made in Section \ref{section:BandComp}.)  For Experiment A, the conservative models do not follow the same qualitative trend as the experimental data while the dissipative models follow the same trend as the experimental data and line up with it well.  For Experiment D, all five models under-predict the experimental data and show no overlap with the experimental data.  This is not unexpected because the NLS-type predictions are corrections to the Stokes predictions and we showed above that the Stokes prediction is inaccurate for Experiment D.  The results for Experiments B and C (plots not included) are qualitatively between the results of Experiments A and D.  For Experiment B, the dissipative models provide predictions of quality similar to Experiment A.  For Experiment C, the dissipative models provide predictions of poor quality, similar to those shown for Experiment D.  We also note that in the conservative models the two indirect methods of prediction do not agree with one another, while in the dissipative models the two indirect methods agree quite well.  Finally, note that Figure \ref{fig:Stokes} was created using the leading-order versions of (\ref{eqn:Stokesb2}) and (\ref{eqn:NLSB2}).  Using the higher-order versions of these equations does not lead to a qualitative (or much of a quantitative) difference in any of the predictions.

Table \ref{table:asymExpt1} shows the percent error norms from the single-mode comparisons between indirect predictions using first harmonic input data and experimental second harmonic data.  The first row in the table, labeled ``Expt,'' corresponds to the difference between the gray dots (experimental measurements) and the black dots (Stokes predictions using (\ref{eqn:Stokesb2})) in Figure \ref{fig:Stokes}.  The other rows in the table correspond to the differences between the gray dots and the curves (predictions obtained from the envelope equations and equation (\ref{eqn:Stokesb2})).  The smallest errors were obtained for Experiment B.  

\begin{table}
\centering
\begin{tabular}{ |c|c|c|c|c| } 
\hline
Equation (\ref{eqn:Stokesb2}) & Expt A & Expt B & Expt C & Expt D \\
\hline
Expt    & $7.38*10^{-1}$ & $7.35*10^{-1}$ & $4.23*10^{0}$ & $3.92*10^{0}$ \\
NLS     & $6.43*10^{0}$ & $4.92*10^{0}$ & $3.28*10^{0}$ & $3.07*10^{0}$ \\
Dysthe  & $1.15*10^{1}$ & $4.08*10^{0}$ & $3.38*10^{0}$ & $2.87*10^{0}$ \\
dNLS    & $1.53*10^{0}$ & $4.25*10^{-1}$ & $3.87*10^{0}$ & $3.59*10^{0}$ \\ 
vDysthe & $1.68*10^{0}$ & $3.25*10^{-1}$ & $3.63*10^{0}$ & $3.52*10^{0}$ \\
dGT     & $1.70*10^{0}$ & $3.05*10^{-1}$ & $3.64*10^{0}$ & $3.52*10^{0}$ \\ 
\hline
\end{tabular}
\caption{Percent error norms, $\mathcal{E}$, using (\ref{eqn:error}) for comparisons between experimental data and the single-mode prediction for $b_2$.}
\label{table:asymExpt1}
\end{table}

\begin{figure}
\centering
  \includegraphics{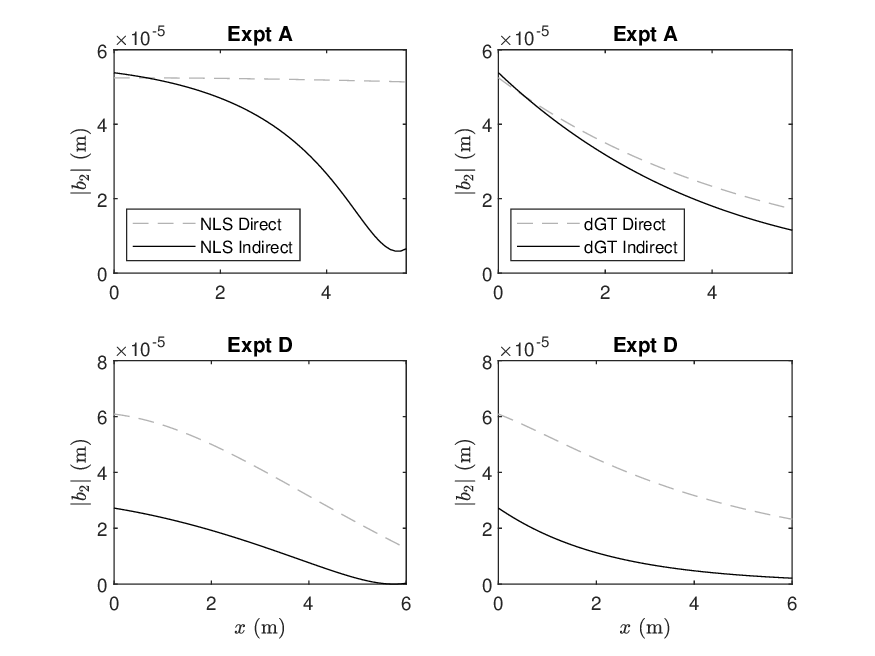}
  \caption{Plots comparing the direct (dashed gray curves) and indirect (solid black curves) for the $6.66$ Hz wave.  The left column shows single-mode comparisons using NLS as the prototypical conservative model. The right column shows single-mode comparisons using dGT as the prototypical dissipative model.}
\label{fig:StokesModel}
\end{figure}

Figure \ref{fig:StokesModel} contains comparisons of the direct (dashed gray curves) and indirect (solid black curves) predictions for the amplitude of the $6.66$ Hz wave in Experiments A and D.  In these plots, NLS serves as a representative conservative model and dGT serves as a representative dissipative model.  The envelope equations predict the evolution of the harmonic bands, but this figure focuses on the 6.66 Hz mode for ease of understanding.  The direct and indirect NLS predictions do not show much overlap in any of the four experiments.  In Experiment A there is only a small difference between the direct and indirect dGT predictions.  This difference is larger in the other three experiments, though they show the same trend.  The Experiment D discrepancy at $x=0$ shows that the second harmonic amplitude is not starting out at the ``correct'' amplitude as far as weakly nonlinear theory is concerned.  The fact that there is less agreement in these comparisons than there was in the comparisons in Figure \ref{fig:Stokes} suggests that the direct and indirect predictions are qualitatively and quantitatively different.

\subsection{Band Comparisons}
\label{section:BandComp}

In this subsection, we test whether the experimental second harmonic bands behave according to the relation between $B_2$ and $B$ given in equation (\ref{eqn:NLSB2}). This comparison includes the 3.33 Hz and 6.66 Hz waves along with their sidebands.  The comparisons in this section are more detailed than those in Section \ref{SMComparisons} because they test the models' ability to predict the evolution of all Fourier amplitudes in the band as opposed to just one Fourier amplitude (the dominant mode).

Plots of the band predictions are not included because there are too many sidebands and because the single-mode prediction plots (see Figure \ref{fig:har2}) show the evolution of the 6.66 Hz wave.  Instead, the information the band prediction provides about the sidebands is examined through the error norm results presented in Table \ref{table:asymExpt2}.  This table shows how accurately the models and equation (\ref{eqn:NLSB2}) predict the evolution of the second harmonic band.  Additionally, just as with the single-mode comparison, there is a quantitative difference between the direct and indirect predictions.

\begin{table}
  \centering
  \begin{tabular}{ |c|c|c|c|c| } 
  \hline
  Equation (\ref{eqn:NLSB2}) & Expt A & Expt B & Expt C & Expt D \\
  \hline
  Expt    & $5.23*10^{-1}$ & $9.31*10^{-1}$ & $9.50*10^{-1}$ & $1.10*10^{0}$ \\
  NLS     & $3.75*10^{-1}$ & $1.11*10^{0}$ & $8.82*10^{-1}$ & $9.82*10^{-1}$ \\
  Dysthe  & $4.42*10^{-1}$ & $1.04*10^{0}$ & $8.96*10^{-1}$ & $9.97*10^{-1}$ \\
  dNLS    & $5.40*10^{-1}$ & $8.96*10^{-1}$ & $9.46*10^{-1}$ & $1.10*10^{0}$ \\ 
  vDysthe & $5.53*10^{-1}$ & $8.91*10^{-1}$ & $9.44*10^{-1}$ & $1.10*10^{0}$ \\
  dGT     & $5.52*10^{-1}$ & $8.87*10^{-1}$ & $9.46*10^{-1}$ & $1.10*10^{0}$ \\
  \hline
  \end{tabular}
  \caption{Percent error norms, $\mathcal{E}$, using (\ref{eqn:error}) for comparisons between experimental data and the band prediction for $B_2$.}
  \label{table:asymExpt2}
  \end{table}

\section{Summary}
\label{SectionSummary}

We have shown that the second harmonic band experienced significantly more dissipative effects than the first harmonic band.  As far as we were able to determine, there is not a simple relationship between the dissipation in these two bands.  We introduced the viscous Dysthe and dissipative Gramstad-Trulsen equations including surface tension.  We showed that the direct comparisons provide predictions for the evolution of the second harmonic band that are more accurate than the indirect comparisons, see the bottom half of Table \ref{table:direct} and Table \ref{table:asymExpt2}.  However, the indirect comparisons provide accurate predictions for some of the experiments.  We showed that the dissipative models (dNLS, vDysthe, and dGT) provided significantly more accurate predictions in almost every comparison we made than did the conservative models (NLS and Dysthe).  The dominant mode in each band was more accurately modeled than the sidebands.  The second harmonic was not as accurately modeled as the first harmonic.  Finally, by comparing the direct and indirect comparisons with each other, we established that these types of predictions are often qualitatively and quantitatively different.  

\section{Acknowledgements}

We thank Camille Zaug, Christopher Ross, and Salvatore Calatola-Young for helpful conversations.  This material is based upon work supported by the National Science Foundation under grants DMS-1716120 (HP, JDC) and DMS-1716159 (DMH).

The datasets generated during and/or analysed during the current study are available in the Harvard Dataverse repository~\cite{DVN/XUB20B2021}.  On behalf of all authors, the corresponding author states that there is no conflict of interest.

\appendix
\section{Spectral Peaks and Means}
\label{AppendixA}
\begin{figure}
  \centering
      \includegraphics[width=16cm]{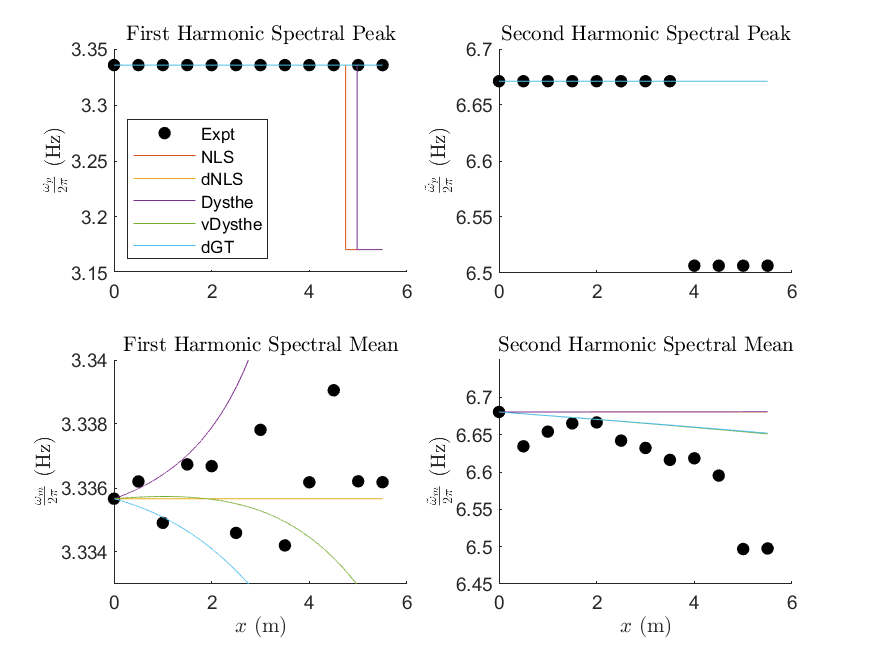}  
      \caption{Plots of the spectral peaks (upper plots) and spectral means (lower plots) for the first harmonic band (left plots) and second harmonic band (right plots) versus distance down the tank for Experiment A.}
      \label{fig:ExptAPeakMean}
\end{figure}

\begin{figure}
  \centering
      \includegraphics[width=16cm]{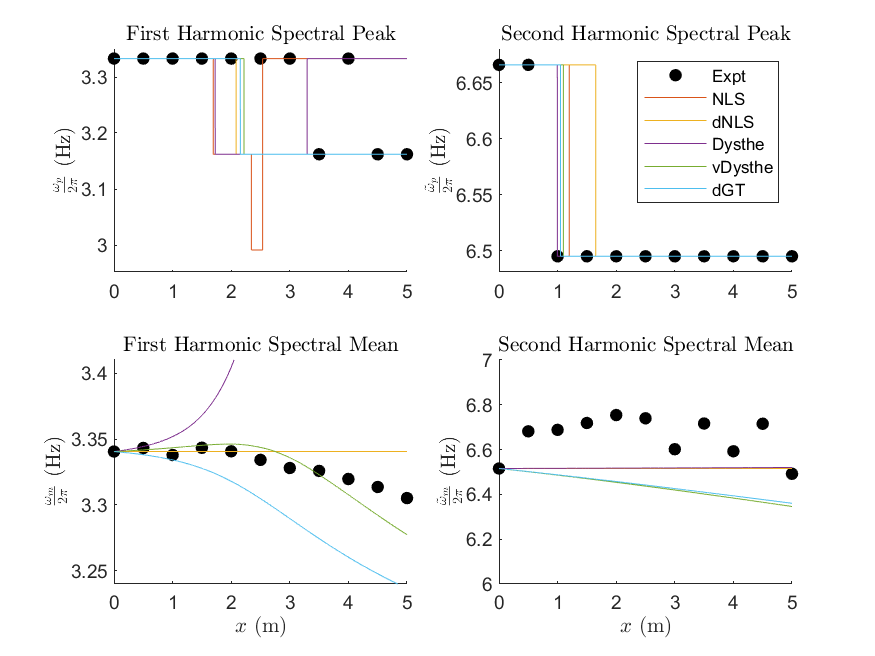}  
      \caption{Plots of the spectral peaks (upper plots) and spectral means (lower plots) for the first harmonic band (left plots) and second harmonic band (right plots) versus distance down the tank for Experiment B.}
      \label{fig:ExptBPeakMean}
\end{figure}

\begin{figure}
  \centering
      \includegraphics[width=16cm]{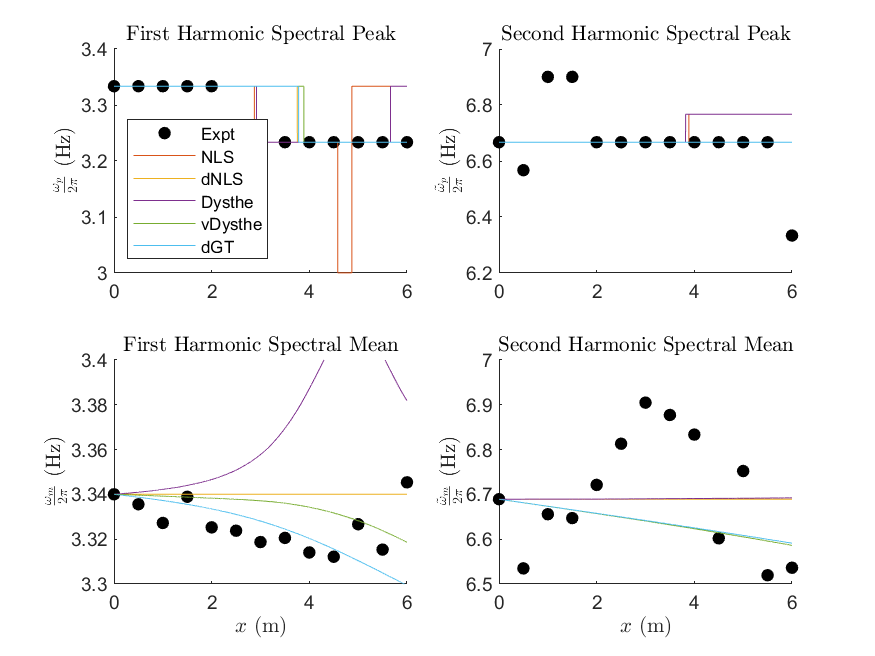}  
      \caption{Plots of the spectral peaks (upper plots) and spectral means (lower plots) for the first harmonic band (left plots) and second harmonic band (right plots) versus distance down the tank for Experiment C.}
      \label{fig:ExptCPeakMean}
\end{figure}

\begin{figure}
  \centering
      \includegraphics[width=16cm]{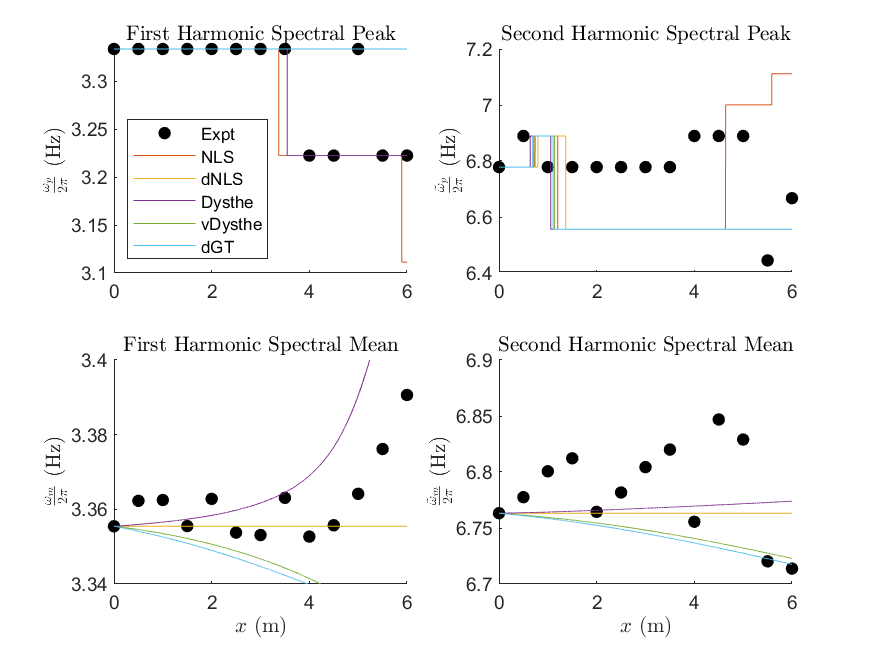}  
      \caption{Plots of the spectral peaks (upper plots) and spectral means (lower plots) for the first harmonic band (left plots) and second harmonic band (right plots) versus distance down the tank for Experiment D.}
      \label{fig:ExptDPeakMean}
\end{figure}

Figures \ref{fig:ExptAPeakMean}-\ref{fig:ExptDPeakMean} contain plots of the evolution of the dimensional spectral peaks and dimensional spectral means for both the first and second harmonic bands for all four experiments.  The spectral peak values were determined by selecting the frequency corresponding to the Fourier mode with largest magnitude at each gauge (for the experiments) or at each $\chi$-step (for the envelope equations).  The (dimensional) spectral mean was computed using
\begin{equation}
  \omega_m=\omega_0+\frac{\mathcal{P}}{\mathcal{M}},
\end{equation}
where the factor of $\omega_0$ is necessary to take into account the fact that NLS-like models factor out the first harmonic, see equation (\ref{eqn:etaexpansion}).  The quantities $\mathcal{P}$ and $\mathcal{M}$ were computed via
\begin{subequations}
  \begin{equation}
    \mathcal{M}=\sum_{j\in \mathbbm{B}}|a_j|^2,
  \end{equation}
  \begin{equation}
    \mathcal{P}=\sum_{j\in \mathbbm{B}}\frac{2\pi j}{L}|a_j|^2,
  \end{equation}
\end{subequations}
where $\mathbbm{B}$ represents the set of frequencies in the band under consideration.  The plots for the spectral peaks and means for the first harmonic band were first shown in Carter et al.~\cite{Butterfield}.  However, the plots shown here are slightly different than those because they include surface tension effects while those in Carter et al.~\cite{Butterfield} did not.

This series of plots show that none of the envelope equations accurately models the evolution of the spectral peaks or means for all four experiments.  This means that none of these models consistently models frequency downshift in the spectral peak or mean sense in an accurate manner. 

\end{document}